\documentclass[a4,11pt]{article}

\usepackage{amsfonts,amsmath,amssymb,bm}
\usepackage{graphicx}
\usepackage{authblk}

\begin{document}


\title{Base spacing distribution analysis for human genome}

\author[2]{Andrzej Z. G\'orski$^1$, Monika Piwowar}

\affil[1]{Institute of Nuclear Physics, Polish Academy of Sciences,
Krak\'ow, Poland}
\affil[2]{Medical College, Jagiellonian University,
Krak\'ow, Poland}

\date{\today}

\maketitle

\begin{abstract}
The distribution of bases spacing in human genome was investigated.
An analysis of the frequency of occurrence in the human genome of different
sequence lengths flanked by one type of nucleotide
was carried out showing that the distribution has no self-similar
(fractal) structure. The results nevertheless revealed several
characteristic features:
(i) the distribution for short-range spacing is quite similar
to the purely stochastic sequences;
(ii) the distribution for long-range spacing essentially deviates
from the random sequence distribution, showing strong long-range correlations;
(iii) the differences between (A, T) and (C, G) bases are quite significant;
(iv) the spacing distribution displays tiny oscillations.
\end{abstract}

\section{Introduction}

The Human Genome (HG) Project was launched in 1990
and was declared complete in 2003.
The reference sequence for the HG was sequenced
across all chromosomes.
Understanding the coding and explanation of the reading
of the genetic information contained
in the full genomic sequence in view ofthe enormity of the
data -- despite analytical efforts -- is still a great challenge
\cite{Green2015,Wilson2015}.
Many studies have proven that the distribution of nucleotides,
as well as whole sequences in the human genome
is not random as it results from the non-random
distribution of coding sequences (genes),  CpG regions,
as well as regulatory, splice and other functional regions
\cite{Denisov2015,Majewsk2002,Louie2003}.
Fragments that do not encode in human DNA also have their distinctive
distribution profile for specific nucleotides \cite{Babarinde2016,Sotero-Cai2017}.

The aim of many investigations has been to pinpoint
important structural characteristics of DNA.
For example, local irregularities along a DNA strand, compared
to surrounding regions, have been associated with biological functionality.
On theother hand, it has been established that the regularity of DNA
recording is characterized, for example, by fragments of introns.
The coding regions in DNA are irregular.
Exon and intron sequences can be identified from trends of the ratio
of the 3-base periodicity to the background noise in the DNA sequences.
Computation of regularities has been also applied to biological weighted
sequences (strings in which a set of letters may occur at each position
with respective probabilities of occurrence) to indicate functionally
significant fragments of DNA.
The above facts indicate that the analysis of nucleotide sequences is still
a big challenge and any advance in describing DNA might provide a valuable
insight.

In this paper the (linear) spacing distribution of each
of four bases in the HG is analysed. We start with the
investigation of possible self-similar (fractal) patterns
and proceed with statistical distribution of the
nearest neighbor spacing for all four bases constituting
the genome.
This type of analysis of data distribution
is widely used not only in physics but also
in other sciences, ranging from bio-medical
\cite{czaszki} to economical \cite{dax} applications.

\section{Materials and Methods}

The Human Genome sequence has been taken from the HG Project in the
FASTA format \cite{dane}. It includes the whole HG
that is about 3~GB large and contains about
2~billions of bases in chromosome's fragments.
The original text file is converted
into numerical files with series of positions of particular
bases,  A, C, G or T, while the other codes were ignored.
The files with concatenated chromosomes are investigated
to reveal averaged global properties of HG and they are the starting
point for further calculations.
It should be stressed that the concatenation has negligible
effect on the results because the number of chromosomes as well
as the largest spacings are of order $10^2$ while the
total length of the HG is of order $10^9$.

\section{Fractal analysis}

First, the possible generalized fractal dimensions
\cite{mandel} of linear distributions of bases A, C, G, T
have been calculated. Such calculations, especially when done with
a software that cannot be fully controlled, can give
misleading results (see e.g. \cite{czaszki,pseudo,box}).
Hence, the calculation has been done with care, using
our own box-counting algorithm code, based on the standard
formula for the generalized fractal dimension \cite{mandel,pseudo}
\begin{equation}
 d_q = {1\over 1-q} \lim_{N\to\infty} {\log \sum_i p_i^q(N)
\over \log N} \equiv {\log Y(N)\over \log N} \ ,
 \label{fracdef}
\end{equation}
where $N$ is the number of (linear) divisions,
parameter $q$ in our case was taken: $q=0, 1, 2$,
for capacity, information and correlation dimensions,
respectively; $p_i(N)$ is number of data points
found in i-$th$ box for a given division $N$.

\begin{figure}[h]
\vspace{30pt}
\begin{center}
  \includegraphics[width=10.0cm,angle=0]{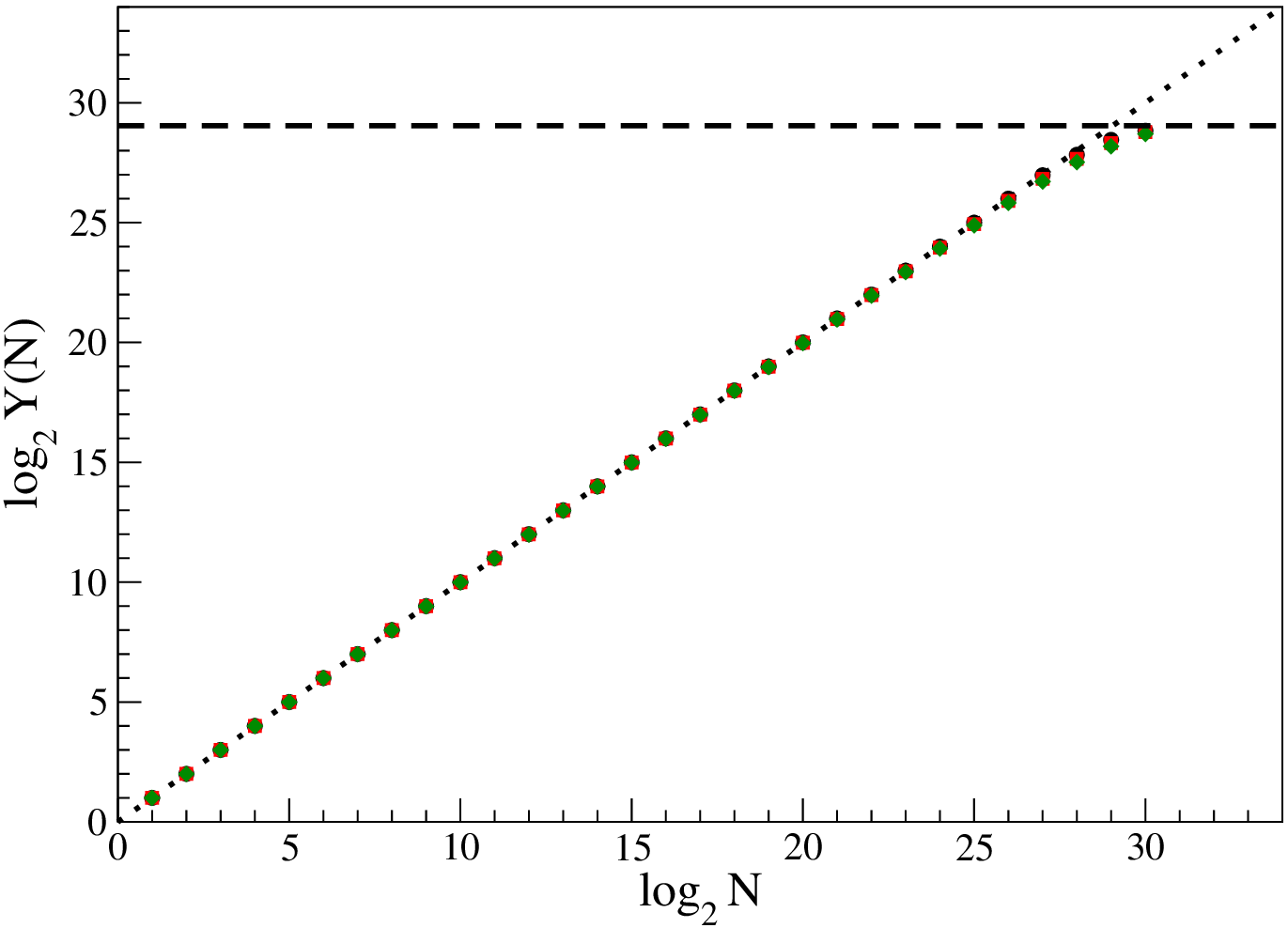}
  \caption{Log-log plot for distribution of base A.
 Circles, squares and diamonds are for capacity ($d_0$),
 information ($d_1$) and correlation ($d_2$) dimension, respectively
 (they strongly overlap).
 The dotted line has the slope coefficient equal 1, like for
 homogeneously or randomly distributed data points. The dashed line
 shows the saturation limit for the ordinate, due to
 the finite size of the data sample.}
\label{fig:fractA}
\end{center}
\end{figure}

  The resulting standard log-log plot used to extract
generalized fractal dimensions for base A is shown in
Fig.~\ref{fig:fractA}. Circles, squares and diamonds are
for capacity ($d_0$),  information ($d_1$) and correlation
($d_2$) dimension, respectively.
In fact, the three symbols can hardly be distinguished,
as they almost perfectly overlap. This excludes multifractality.
Moreover, they are placed along the dotted line that 
has the slope coefficient equal $1.00$,
like for homogeneously or randomly distributed data points.
The dashed line shows the saturation limit for the ordinate,
$\log_2(n_{dp})$, where $n_{dp}$ is the total number of data
points due to the finite size of the sample \cite{pseudo}.
Fig.~\ref{fig:fractA} gives results for the base A, only.
However, identical plots were obtained for all four bases,
as well as for selected single chromosomes.
Moreover, almost identical plots were obtained for randomly
generated data samples of the same size.

Fig.~\ref{fig:fractA} implies that the data set has 
integer (non-fractal) dimension precisely equal
to 1.00.
Clearly, due to the Hentschel-Procaccia inequality\cite{HP}
$d_q = 1.00$ for all $q<2$, as the function $d_q)$ is
monotonic.
Calculations for higher values of $q$ were not performed
because for very small $p_i(N)$ in sum in eq.~\ref{fracdef}
their high powers are beyond any reasonable compiler accuracy.
Hence, one has to conclude that the spacing distribution of bases
in HG does not show any trace of direct self-similarity,
fractal or multifractal structure.

In this place, it is worth to remind, that within the
2-dimensional Chaos Game Representation (CGR) of DNA
sequences \cite{cgr} their fractal structure is well
established by many authors (see e.g. \cite{bmc}).
Self-similarity in those cases is due to
the special properties of the CGR transformation,
that is a kind of recurrence plot technique \cite{rplot}.
These techniques are useful as randomness tests for
random number generators \cite{cgr}, as well as stationarity
tests for time series \cite{dax}.
However, they do not imply self-similarity of the data sample
by itself.

Even though the investigated data samples are not self-similar,
and they were shown to have high entropy \cite{entr} ---
like random sequences --- they are definitely not purely random.
This will be shown in the following section.
Moreover, even a highly structured data can resemble random
series after compression, as the data compression algorithms
increase the Shannon entropy.

\section{Spacing distribution analysis}

\begin{figure}[h]
\vspace{30pt}
  \centering
  \begin{minipage}[b]{0.47\textwidth}
    \includegraphics[width=\textwidth]{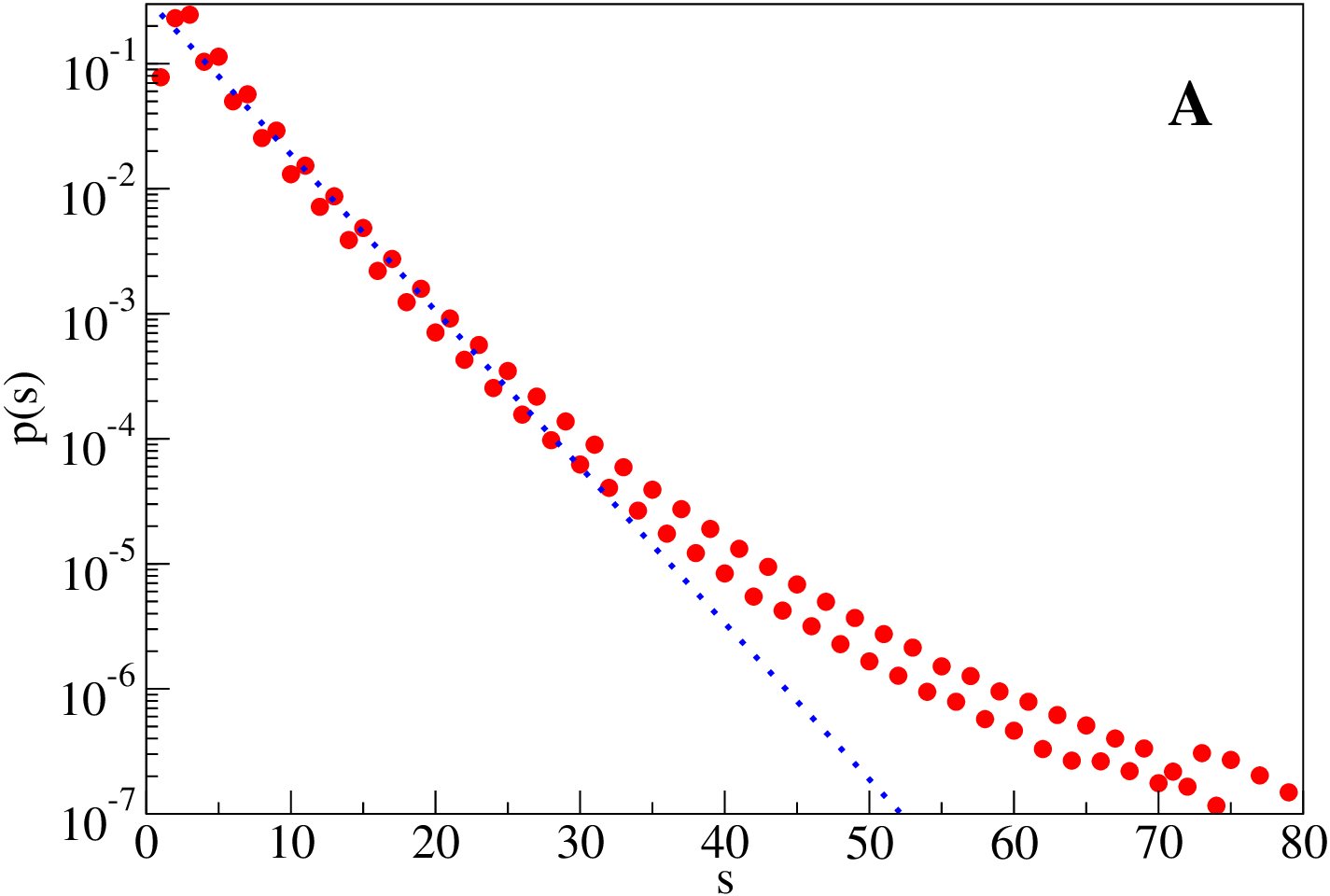}
  \end{minipage}
  \hfill
  \begin{minipage}[b]{0.47\textwidth}
    \includegraphics[width=\textwidth]{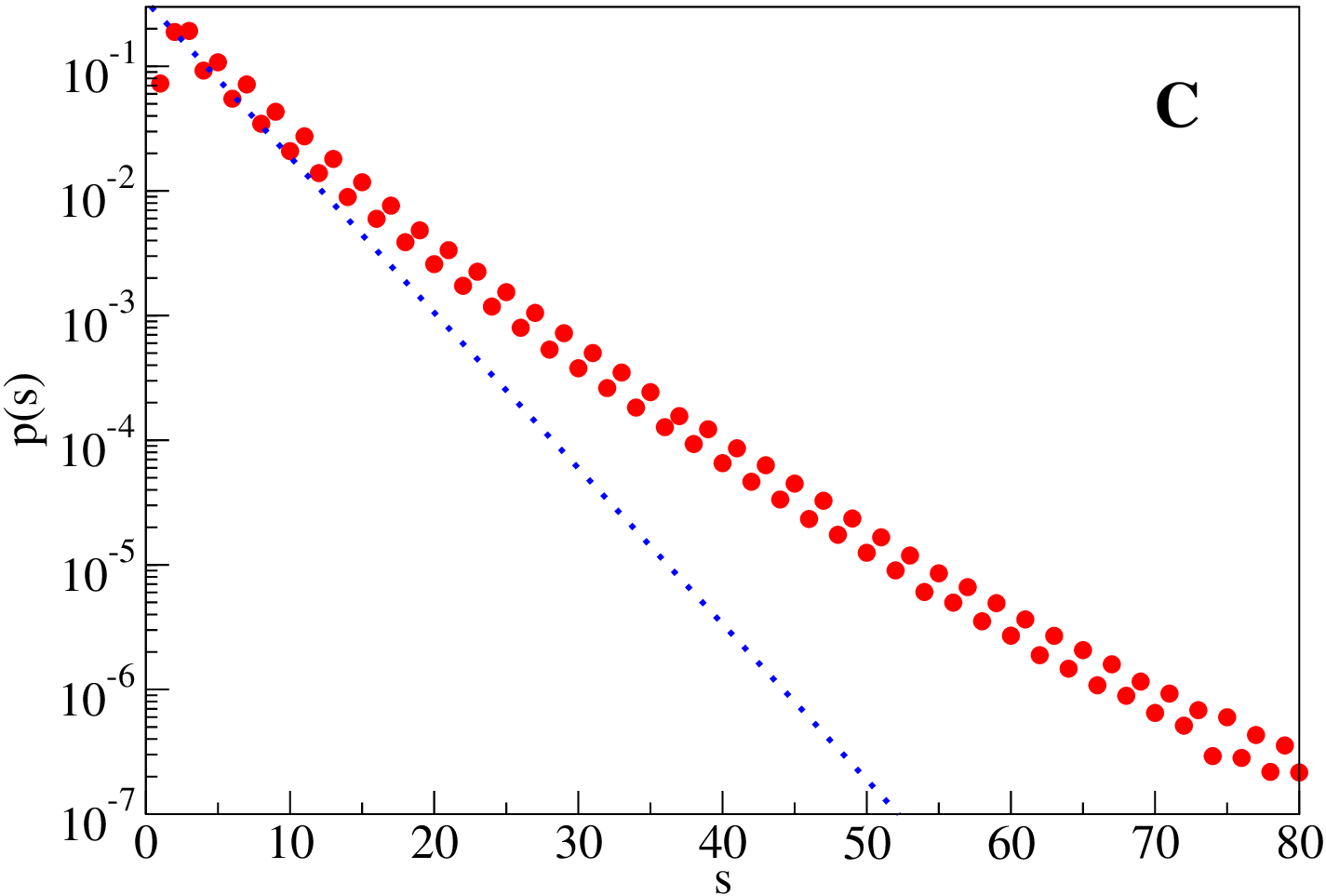}
  \end{minipage}
  \caption{Normalized histogram of spacing distribution, $p(s)$,
           for bases A and C.
           The dotted line corresponds to a purely random
           distribution. Horizontal axis gives spacing distance
           and the vertical axis gives probability.}
  \label{fig:hist1}
\end{figure}

\begin{figure}[h]
\vspace{30pt}
  \centering
  \begin{minipage}[b]{0.47\textwidth}
    \includegraphics[width=\textwidth]{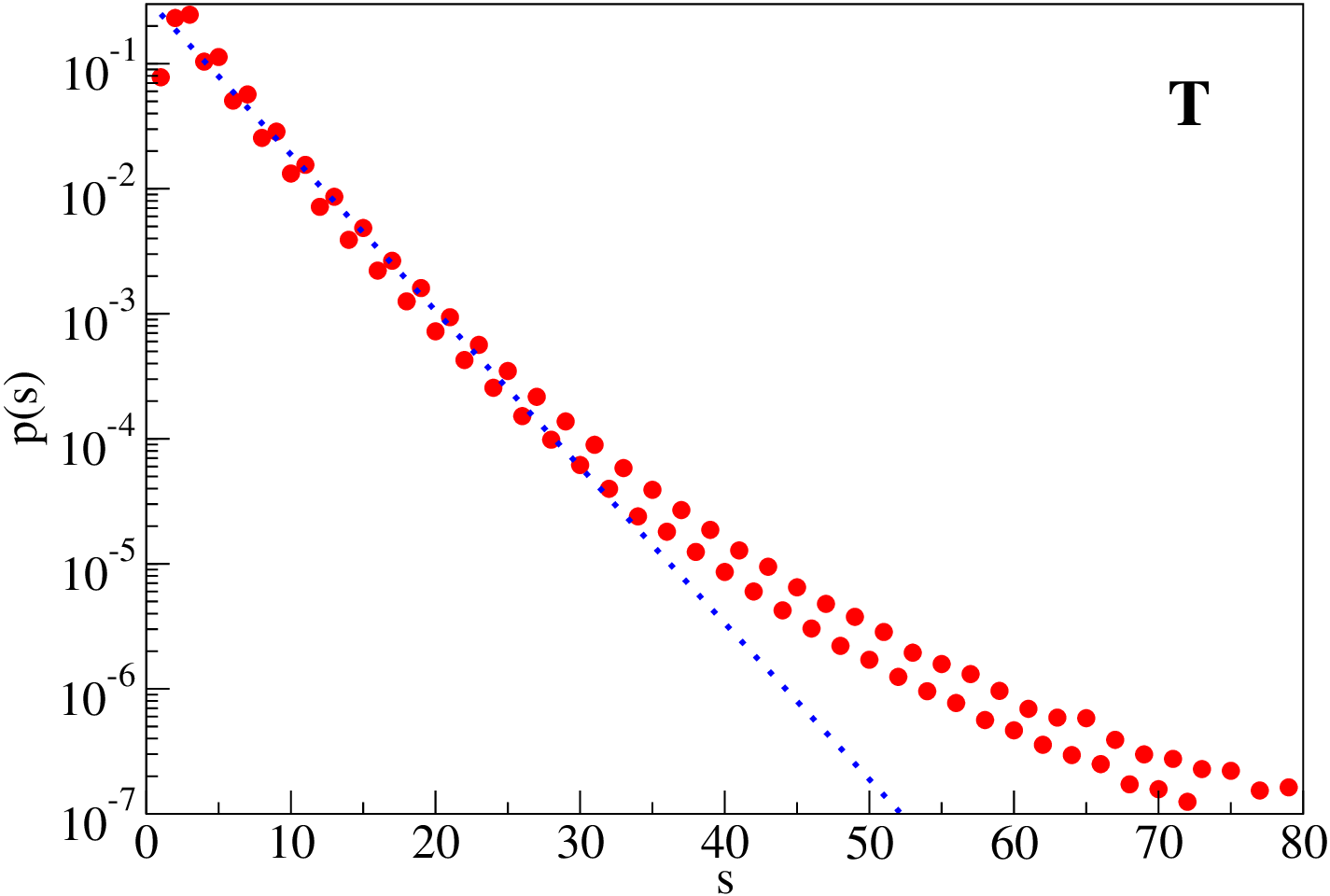}
  \end{minipage}
  \hfill
  \begin{minipage}[b]{0.47\textwidth}
    \includegraphics[width=\textwidth]{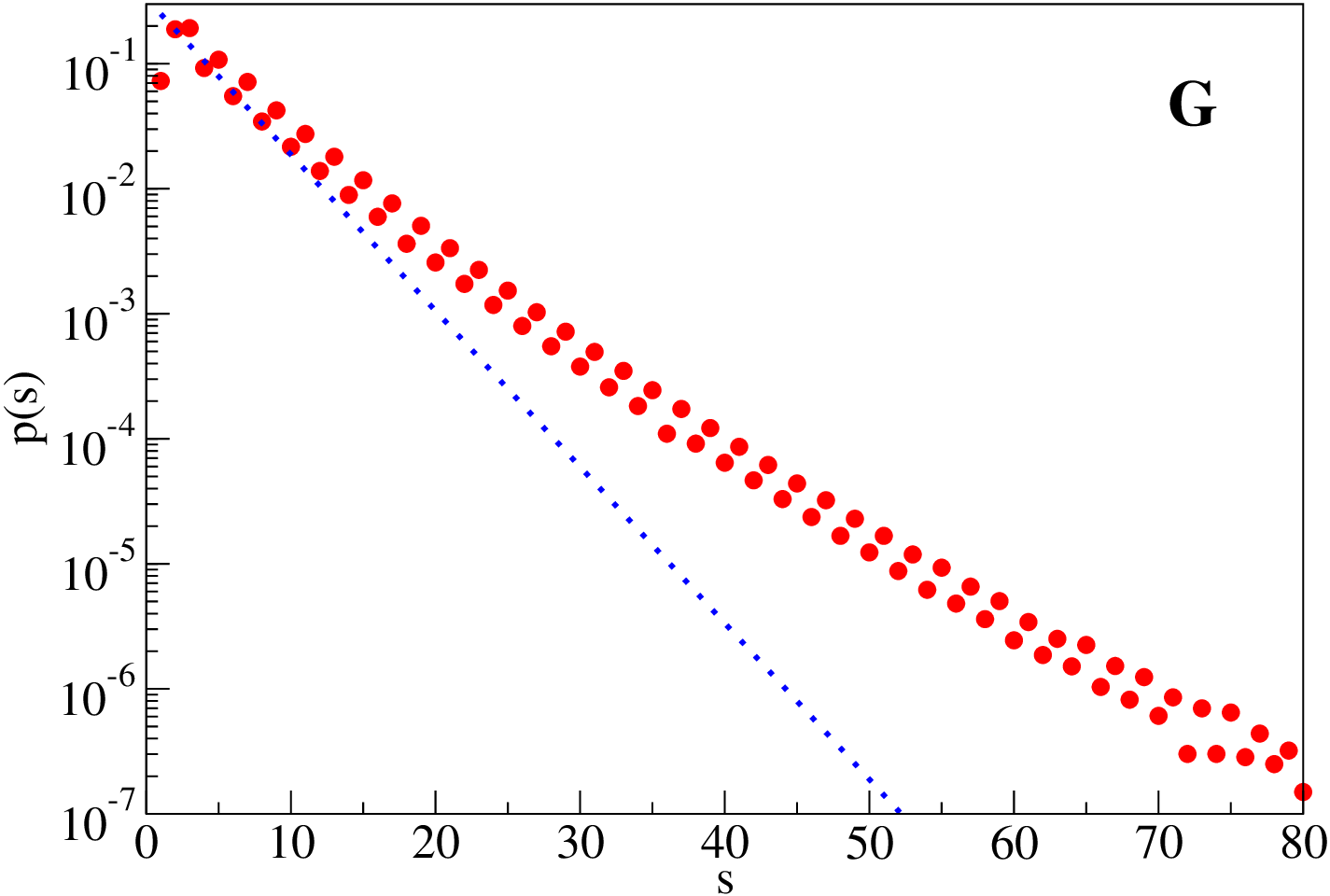}
  \end{minipage}
  \caption{Normalized histogram of spacing distribution, $p(s)$,
           for bases T and G.
           The dotted line corresponds to a purely random
           distribution. Horizontal axis gives spacing distance
           and the vertical axis gives probability.}
  \label{fig:hist1b}
\end{figure}

 In this section we analyze the spacing distribution,
$p(s)$, between bases of the same type. Here, spacing ($s$)
is defined as the distance between two closest neighbors
of the same type. For example, for the base A
and the sequence AA the spacing of bases A is $s=1$.
For the sequence AXA, where X is any base except A,
the spacing is $s=2$ {\it etc}.
In Figs.~\ref{fig:hist1},\ref{fig:hist1b} the circles show (normalized)
probabilities, $p(s)$, of a given spacing in the sample.
In addition, we added a dotted line that corresponds to the
uniform random distribution of bases,
\begin{equation}
 p_{rand}(s) = 1/3 \times (3/4)^s\ ,
\quad {\rm where} \ \sum_{s=1}^{s=\infty} p(s) = 1 \ .
 \label{randistr}
\end{equation}

 For the HG data the spacing distribution has cutoff for $s_{max}$
that is at most of order $10^3$.
The total number of occurrences of base A (and T)
is about $5.5\times 10^8$ and for base C (and G)
about $4.1\times 10^8$.
 In Fig.~\ref{fig:hist1} plots are given for bases A and C,
while in Fig.~\ref{fig:hist1b} for bases T and G.
Both pairs of plots are similar, in accordance with the
Chargaff's rule.
All probability distributions are normalized to unity to
enable comparison of samples with different sizes.

\begin{figure}[h]
\vspace{30pt}
  \centering
  \begin{minipage}[b]{0.47\textwidth}
    \includegraphics[width=\textwidth]{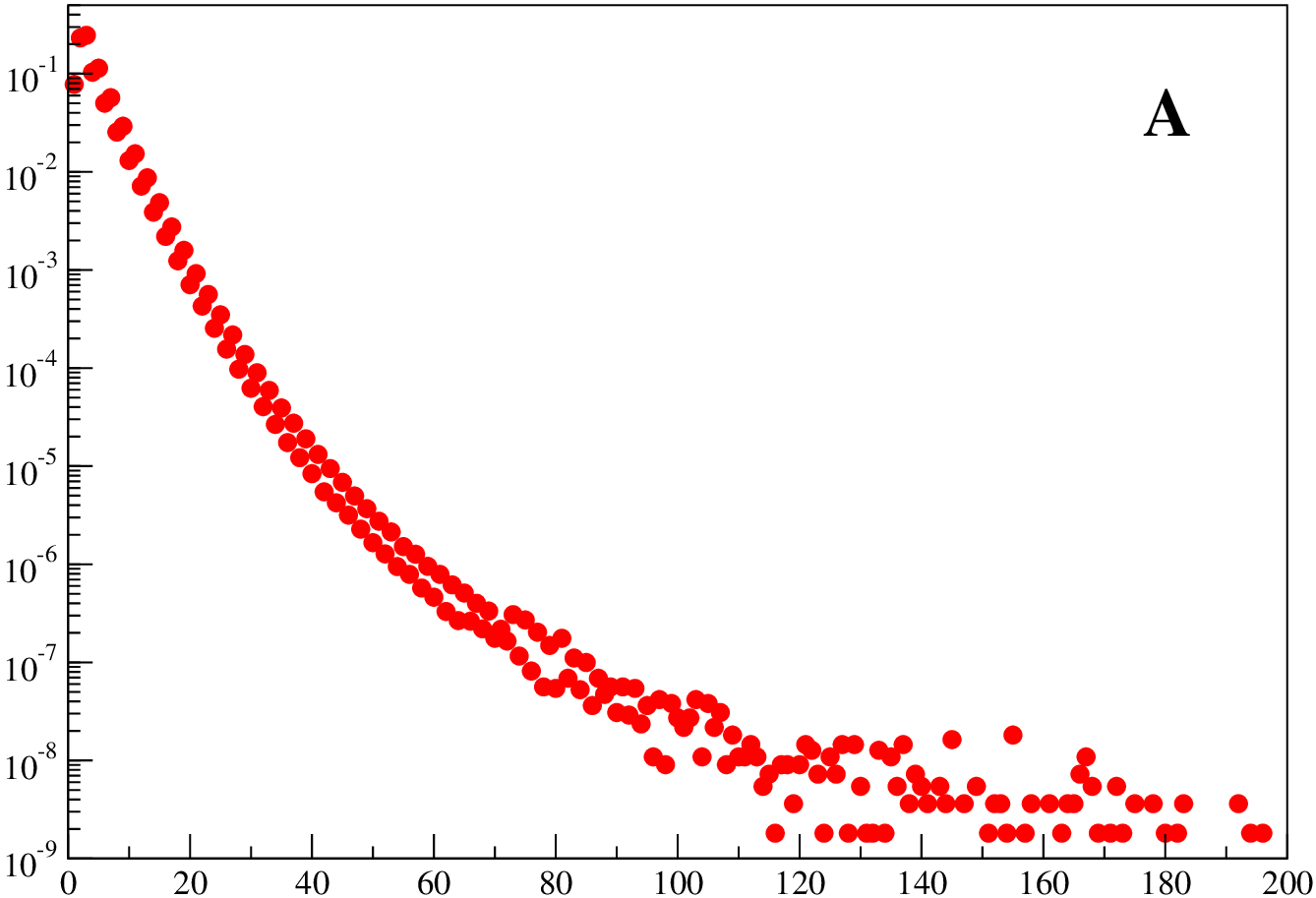}
  \end{minipage}
  \hfill
  \begin{minipage}[b]{0.47\textwidth}
    \includegraphics[width=\textwidth]{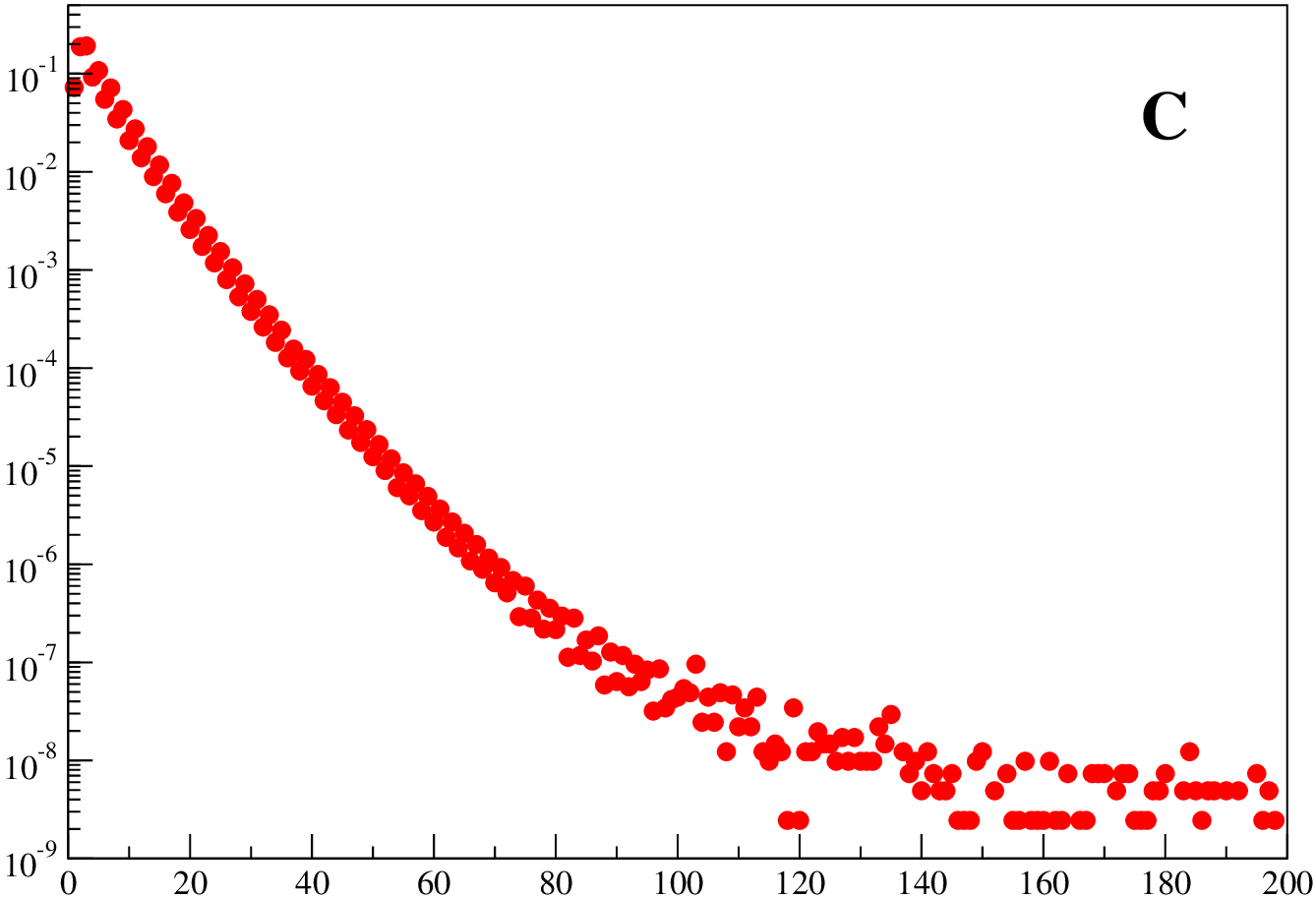}
  \end{minipage}
  \caption{Normalized histogram of spacing distribution for bases
           A and C with tail up to $s=200$.
           Horizontal axis gives spacing distance
           and the vertical axis gives probability.}
  \label{fig:hist2}
\end{figure}

\begin{figure}[h]
\vspace{30pt}
  \centering
  \begin{minipage}[b]{0.47\textwidth}
    \includegraphics[width=\textwidth]{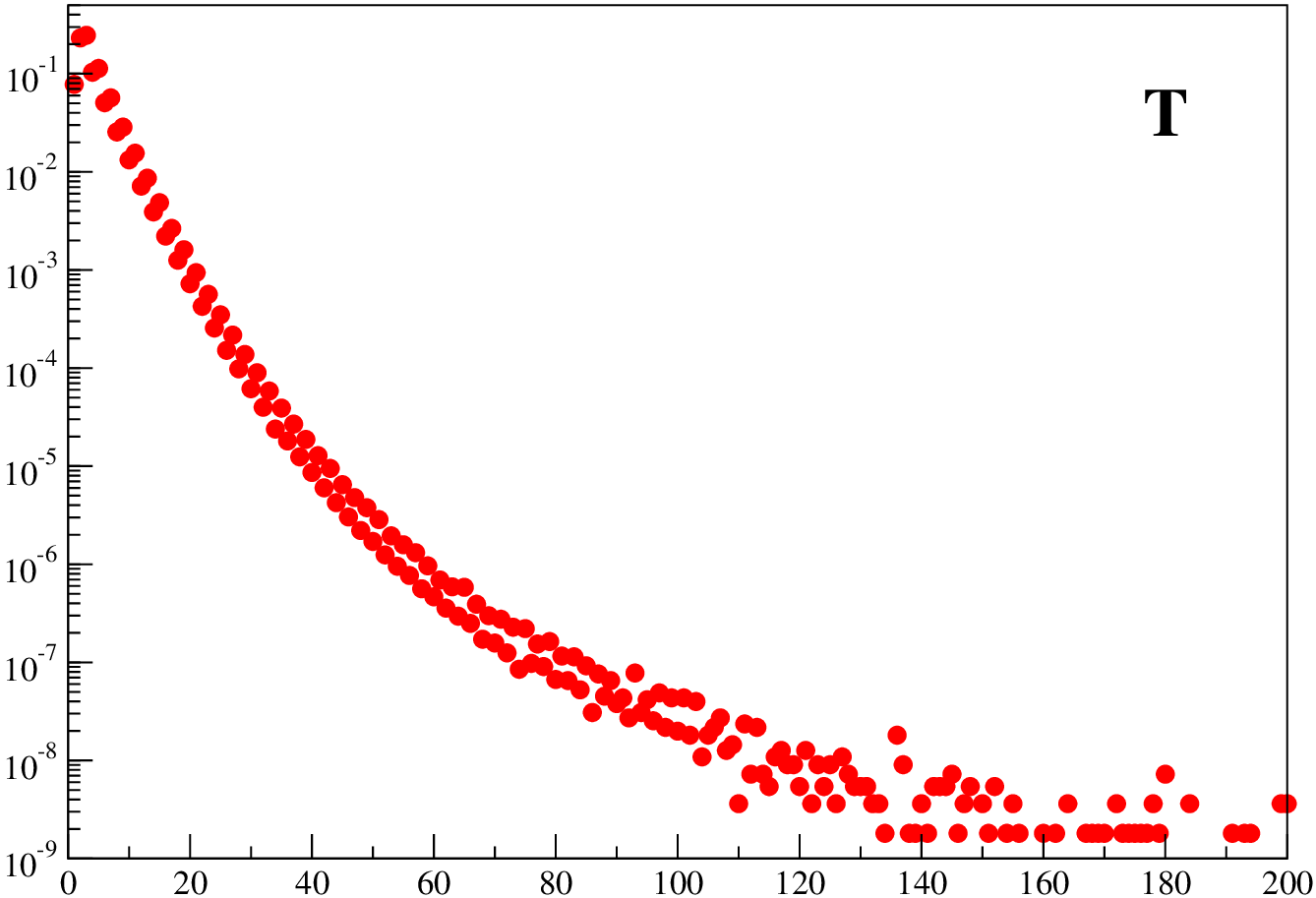}
  \end{minipage}
  \hfill
  \begin{minipage}[b]{0.47\textwidth}
    \includegraphics[width=\textwidth]{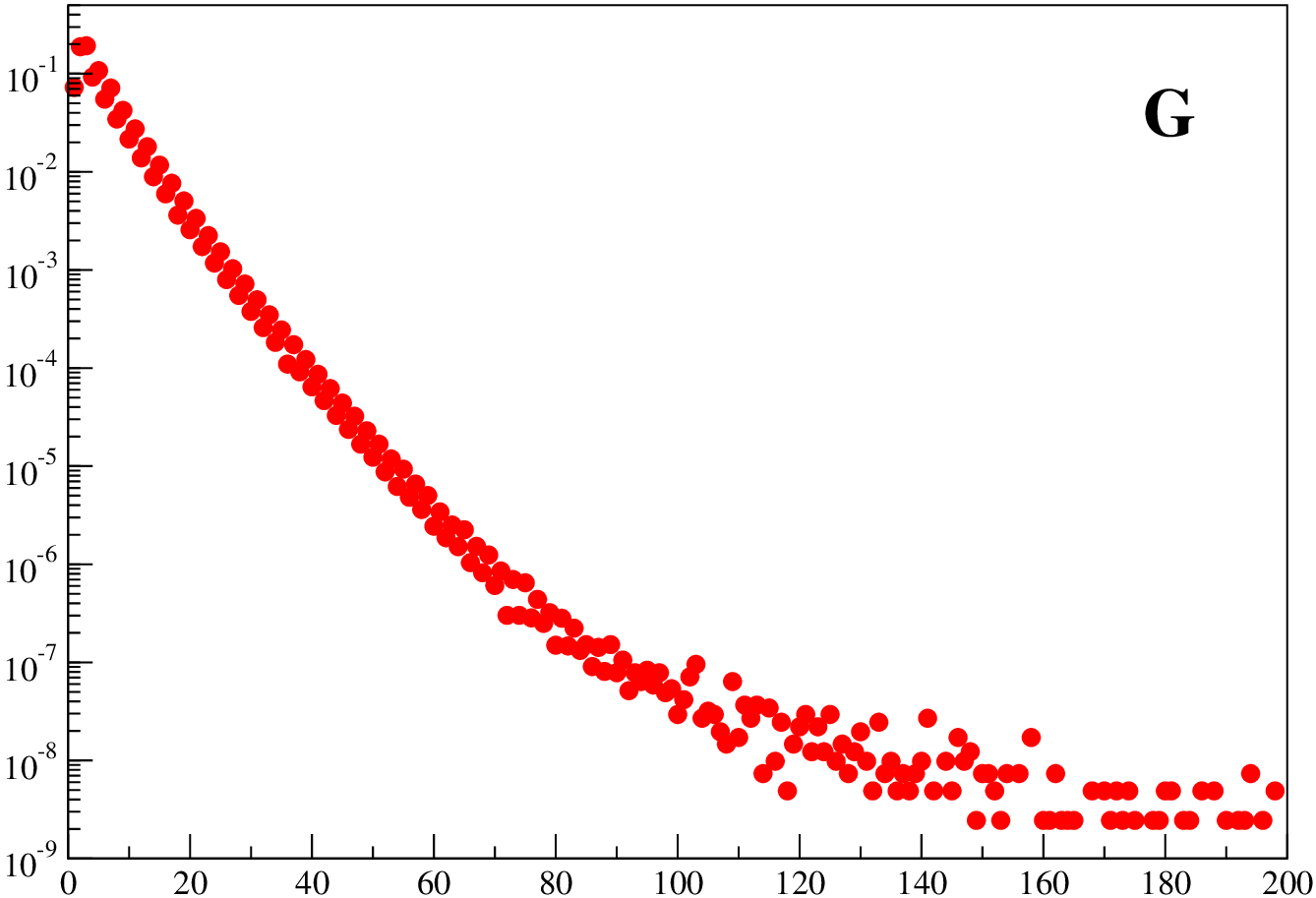}
  \end{minipage}
  \caption{Normalized histogram of spacing distribution for bases
           T and G with tail up to $s=200$.
           Horizontal axis gives spacing distance
           and the vertical axis gives probability.}
  \label{fig:hist2b}
\end{figure}

 In Figs.~\ref{fig:hist2} and \ref{fig:hist2b} the tails
of the histograms are shown up to $s=200$. Here, one can see
that for larger spacings ($s$) the tail is getting fat and
strongly deviates from exponential behavior.
Also, one can see a kind of phase transition at $s_2\approx 80$
and the histograms' bins are more randomly distributed.
For $p(s)$ approaching $10^{-9}$ there are only single data points
per bin and the statistics becomes less reliable. Hence, though
the single events are up to $s\approx 1000$ they are not displayed.
It should be stressed that fat tails are also common for self-organizing
systems in economy, sociology {\it etc.}, where long-range
correlations (LRC) occur \cite{dax}.

 Closer examination of spacing distributions reveals
several characteristic features that are listed below:

\noindent (i) For small spacing (about $s_1^A\approx 30$ for A and T
bases, but $s_1^C<10$ for C and G bases) the distributions are
quite close to the purely random distribution.
However, for larger spacings the distributions strongly deviate
form randomness.

\noindent (ii) The long tails of the distributions are strongly
enhanced ('fat tails') in comparison with the random distribution.
This suggests strong long distance correlations.

\noindent (iii) In general, behavior of base A is similar as for
base T, and the same holds for the (C,G) pair, though both pairs
behave in different way. This can be viewed as another manifestation
of the Chargraff's rule.

\noindent (iv) For odd spacing ($ s=3,5,7,...$) probability is
higher than for their even predecessors. And the difference
is slightly higher for (C,G) bases than for (A,T) bases.
This is a kind of small high frequency oscillations
in the distributions.

\section{Discussion and Conclusion}

It has been shown that the base spacing distribution
in the HG is not random, though its high entropy.
This is confirmed by the known fact that the nucleotide
composition of the DNA sequence determines its spatial
structure, function and stability of the spatial structure
of the nucleic acid \cite{Vologodskii2013,Vologodskii2018,Traversi2005}

It has been found that the analyzed distribution has no fractal
structure and for small spacings ($s<s_1$) it is close to random
distribution (exponential decay).
Analogous conclusion that the so-called random matches always dominate
the distribution for small lengths has also been found recently
for eukaryotic genomes \cite{Massip2015}, with similar suggested
estimate, $s_1 \approx 25$.

 On the other hand, for larger spacing the distribution shows
strong correlations and fat tails.  Existence of LRC within the genome
of living organism has immense importance in understanding
the language of DNA sequences. However, the biological meaning of
the LRC in DNA is, as yet, not clear.  It is still an open and challenging
problem.
LRC has been suggested to be related to the duplication of DNA fragments
Some authors claim that LRC occur only on intron containing
DNA sequences, some however, that LRC do not distinguish between the intron
and intronless DNA sequences.
There have also been reports that LRC can be related to the nucleosomal
structure and dynamics of the chromatin fiber.
 Our results are in agreement with conclusions reached by other
authors, see {\it e.g.} \cite{Massip2015,Messer2007}.
Moreover, the LRC have been shown important to the persistence of resonances
of finite segments \cite{Albuq2005}.

 In addition, for large distances, $s>s_2\approx 80$, strong
variability around any smooth interpolation was found.
Variability of long nucleotide fragments is most likely responsible
for structural variation, which is read by molecules interacting
with DNA, which are conformationally sensitive.
Attempts are made to analyze the variability of the DNA sequence
in terms of structural variation resulting from variation at
the sequence level by e.g. parametric and non-parametric entropy measures.
 Also, one can speculate, that relatively high entropy of the sequences
reported previously \cite{entr} (and some similarity to random series)
may be an effect of a kind of data compression algorithm.

 Finally, the A-T and C-G bases have very similar distributions
that is in accordance with the Chargaff's rule.
On the other hand, there is clear difference between the two
pairs. The C-G bases have significantly higher probability for
larger spacing (fatter tails).
For $s=50$ the probability for C is about $10$ times higher.
On the other hand, the tail for C is shorter and its
maximum is slightly higher. Such behavior have also been found
for genomes of other spacies \cite{Afreixo2009}.


\begin{thebibliography}{99}
\begin{small}
\itemsep-2pt


\bibitem{Green2015}
 E.D Green, J.D. Watson, F.S Collins,
"Human Genome Project: Twenty-five years of big biology",
Nature. 526 (2015) : 29–31.

\bibitem{Wilson2015}
B. Wilson, S.G. Nicholls,
"The Human Genome Project, and recent advances in personalized genomics",
Risk Manag Healthc Policy. 8 (2015) : 9.

\bibitem{Denisov2015}
S. Denisov, G. Bazykin, A. Favorov, A. Mironov, M. Gelfand,
"Correlated Evolution of Nucleotide Positions within Splice Sites in Mammals",
 PLoS One. 10(12) (2015) :e0144388

\bibitem{Majewsk2002}
J. Majewski, J. Ott,
"Distribution and characterization of regulatory elements in the human genome",
Genome Res. 12(12) (2002) :1827-36

\bibitem{Louie2003}
E. Louie, J. Ott,J.  Majewski,
"Nucleotide frequency variation across human genes",
Genome Res. 13(12) (2003) :2594-601

\bibitem{Babarinde2016}
I.A. Babarinde, N. Saitou,
"Genomic Locations of Conserved Noncoding Sequences and Their
Proximal Protein-Coding Genes in Mammalian Expression Dynamics",
Mol Biol Evol. ;33(7) (2016) :1807-17.

\bibitem{Sotero-Cai2017}
C.G. Sotero-Caio, R,N  Platt RN, A. Suh, D.A. Ray,
Evolution and Diversity of Transposable Elements in Vertebrate Genomes.
Genome Biol Evol. 1;9(1) (2017) :161-177.

\bibitem{czaszki}
A.Z. G\'orski, J. Skrzat,
"Error estimation of the fractal dimension measurements of cranial sutures",
J. Anat. {\bf 208} (2006) 353-359.

\bibitem{dax}
A.Z. G\'orski, S. Dro\.zd\.z, J. Speth,
"Financial multifractality and its subtleties: an example of DAX",
Physica A 316 (2002) 496-510.

\bibitem{dane}
{\tt www.ncbi.nlm.nih.gov},
Genome Reference Consortium, Human Reference 38,
FASTA file, p12, release Dec 27, 2017.

\bibitem{mandel} B.B. Mandelbrot, "The Fractal Geometry of Nature"
San Francisco, Freeman 1982.

\bibitem{pseudo}
A.Z. G\'orski,
"Pseudofractals and the box counting algorithm",
J. Phys. (GB) {\bf A34} (2001) 79337940.

\bibitem{box}
A.Z. G\'orski, M. Str\'o\.z, P. O\'swi\c ecimka, J. Skrzat,
"Accuracy of the box-counting algorithm for noisy fractals",
Int. J. Mod. Phys. {\bf 27} (2016) 1650112-1-12.

\bibitem{HP}
H.G.E. Hentschel, I. Procaccia,
"The infinite number of generalized dimensions of fractals and
strange attractors",
Physica D 8 (1983)  435-444.

\bibitem{cgr}
H.J. Jeffrey,
"Chaos game representation of gene structure",
Nucleic Acids Res. 18 (1990) 2163-2170.

\bibitem{bmc}
P.A. Moreno, et al.,
"The human genome: a multifractal analysis",
BMC Genomics 12 (2011) 506 1-17.

\bibitem{rplot}
J.-P. Eckmann, S.O. Kamphorst, D. Ruelle,
"Recurrence plots of dynamical systems",
Europhys. Lett., 5(9) (1987) 973-977.

\bibitem{entr}
A.O. Schnitt, H.P. Herzel,
"Estimating the Entropy of DNA Sequences",
J. Theor. Biol.  1888 (1997) 369-377.

\bibitem{Vologodskii2013}
A.D. Vologodskii, M. Frank-Kamenetskii,
"Strong bending of the DNA double helix"
Nucleic Acids Res. 41 (2013) : 6785–6792

\bibitem{Vologodskii2018}
A. Vologodskii, M.D.  Frank-Kamenetskii,
"DNA melting and energetics of the double helix",
Phys Life Rev.  25 (2018) :1-21

\bibitem{Traversi2005}
A. Travers,
"DNA dynamics: bubble “n” flip for DNA cyclisation?",
Curr Biol. 15 (2015) : R377-9

\bibitem{Massip2015}
F. Massip, M. Sheinman, S. Schbath, P.F. Arndt,
"How evolution of genomes is reflected in exact DNA match statistics",
Mol. Biol. Evol. 32 (2015) 524-535.

\bibitem{Messer2007}
P.W. Messer, R. Bundschuh, M. Vingron, P.F. Arndt,
"Effects of long-range correlations in DNA sequence
aligment score statistics",
J. Comput. Biol. 14 (2007) 655-668.

\bibitem{Albuq2005}
E.L. Albuquerque, M.S. Vasconcelos, M.L. Lyra, F.A. de Moura,
"Nucleotide correlations and electronic transport of DNA sequences",
Phys. Rev. E 71 (2005) 021910.

\bibitem{Afreixo2009}
V. Afreixo, C.A. Bastos, A.J. Pinho, S.P. Garcia, P.J. Ferreira,
"Genome analysis with inter-nucleotide distances",
Bioinformatics 1 (2009) 3064-3070.


\end{small}

\end{thebibliography}
\end{document}